


\magnification = 1200 \baselineskip = 18pt

\centerline{\bf Dynamic Fractional Stark Ladders in DC--AC Fields} \vskip
.3in \centerline {X. -G. Zhao, R. Jahnke$^*$ and Q. Niu$^*$} \vskip .2 in
\centerline {CCAST (World Laboratory) P.O.Box 8730, Beijing 100080}
\centerline {and Institute of Theoretical Physics, Academia Sinica,}
\centerline{ P.O.Box 2735, Beijing 100080, China} \vskip .5 in
\centerline{$^*$Department of Physics, The University of Texas at Austin,}
\centerline {Austin, Texas 78712--1081} \vskip .5in

\centerline { \bf
Abstract} A single band in a spatially periodic system splits into a series
of quasienergy subbands under the action of DC--AC electric fields. These
dynamic fractional Stark ladders should be observable in an investigation
of optical absorption.

\vskip .3in PACS$:$ 72.10.Bg, 73.20.Dx \vfill\eject

In the early sixties,
Wannier [1] showed that Bloch electrons in a constant electric field form
equally
spaced discrete levels, the so called Wannier--Stark ladders (WSL), with
localized wave functions. This point of view was later shown
not to be strictly correct [2], but  valid when Zener tunneling
could be neglected [3].  The experimental observation of the WSL in solids
suffers
certain difficulties because the required electric field strength is too  high
[4]. With the
advent of semiconductor superlattices this obstacle has been eliminated since
the miniband
Brillouin zone, and therefore the required electric field strength, for
such structures is dramatically reduced as a result of the large
periodicity [5]. A method for detecting the WSL was proposed by Bleuse,
Bastard, and Voisin based on a tight--binding calculation of the optical
absorption coefficient [6].
To date, a number of experiments confirming the existence of
the WSL have been carried out [7].

The situation where a Bloch electron moves
under the action of an AC electric field has recently been
studied. It is found that an initially localized electron will remain
localized if the ratio of the field magnitude  and the field frequency is a
root of the ordinary Bessel function of order zero [8]. This phenomenon
involving the dynamic localization of moving carriers was reexamined more
recently in a study of quasienergy minibands in superlattices [9]. There it
is demonstrated that the occurence of band collapse coincides with the onset
of dynamic localization.

A natural next step is to inquire what happens when a Bloch electron
is exposed simultaneously to both DC and AC fields.  When
the DC field is such that its WSL frequency is an integer multiple
of the AC frequency, the quasienergy spectrum of the pure AC case has
been found qualitatively unchanged [10].  When the ratio between the two
frequencies,
to be called the {\it matching ratio}, is
not an integer, the competition between the two fields should give
rise to rich physical phenomena.  For instance, it has been demonstrated
that when the matching ratio is an irrational number, the electrons
remain localized as in a Stark ladder state.  However, when the matching ratio
becomes
a rational number, the electrons can become delocalized as in the general
case of a pure AC field.
The experimental observation of this new dynamic localization/delocalization
phenomenon
should not be, in contrast to the purely AC field case, too difficult [11].
This paper
is chiefly concerned with a systematic study of the quasienergy spectrum for
rational
matching ratios of DC--AC systems of the type discussed above.

In this paper, we will derive analytically the quasienergy spectrum in a
simple model, and show how the parent band is split into several
quasienergy subbands with a fractional ladder structure, which we call, in
contrast to the integral Stark ladder [1], the dynamic fractional Stark ladder.
The quasienergy spectrum will then be ploted for a range of rational values,
$p/q$,
of the matching ratio. A
perturbative calculation of optical absorption will show how this fractional
ladder
structure should be observable.

For the sake of simplicity, we focus on an electron in a one dimensional
system with a spatially periodic potential under the influence of a
temporally periodic electric field. The Hamiltonian can be written as
$$H(x,t) = H_{0}(x) - xF(t), \eqno(1)$$ where $H_{0}(x)$ is the field free
part of the Hamiltonian and is spatially periodic with period 1, and $F(t)$
is the force on the electron due to the external field, and is periodic in
time with period $T$. It then follows that the Hamiltonian (1) is also
periodic in time,  $H(x,t+T) = H(t)$. Hence, Floquet's theorem asserts that
the wave functions of $H(x,t)$ can be expressed as  $\psi(x,t) =
\exp(-i\epsilon t)u_{\epsilon}(x,t) $ with $u_{\epsilon}(x,t+T) =
u_{\epsilon}(x,t)$, where $\epsilon$ is the quasienergy [9] and
$u_{\epsilon}(x,t)$ satisfies the Schr\"odinger equation
$$[H(x,t)-i{\partial\over\partial t}]u_{\epsilon}(x,t)=
\epsilon u_{\epsilon}(x,t).  \eqno(2)$$
Since this equation contains only the first order time derivative, the general
periodic condition may be conveniently replaced by $u_{\epsilon}(x,T) =
u_{\epsilon}(x,0)$.  (For a second order differential equation, the first order
time derivatives of the solution at $t=0$ and $t=T$ are also required to be the
same.)

						Assuming that Zener
tunneling can be neglected, we only pay attention to the single--band case,
writing $u_{\epsilon}(x,t)= \sum_{m}u_{\epsilon m}(t) \phi(x-m)$,
where $\phi(x-m)$ is the single-band Wannier function associated
with site $m$. The coefficients $u_{\epsilon m}(t)$ are periodic in time
and satisfy the evolution equations
$$i{d \over
dt}u_{\epsilon m}(t)
=\sum_{m'}\langle0\vert{H_0}\vert{m'}\rangle
u_{\epsilon (m+m')}(t)-[\epsilon+mF(t)]u_{\epsilon m}(t)
\eqno(3)$$
with $\langle0\vert{H_0}\vert{m}\rangle$ being the matrix element between two
Wannier
functions.  By introducing
$$C_m(t)=e^{-i[\epsilon t+m\theta(t)]}u_{\epsilon m}(t), \eqno(4)$$
where $-\theta (t)$ is the vector potential defined by
$\theta(t)=\int_0^t dt' F(t')$,
we obtain
$$i{d \over
dt}C_{ m}(t)
=\sum_{m'}\langle0\vert{H_0}\vert{m'}\rangle e^{i m'\theta(t)}
C_{m+m'}(t).
\eqno(5)$$
Employing the discrete Fourier transform $C_{k}(t) =
\sum_{m}{C_{m}(t)\exp(-imk), (-\pi\le{k}<\pi)}$ leads to the solution of
(5) as
$$C_{k}(t) =
C_{k}(0)\exp\{-i\int_{0}^{t}{dt^\prime}{E(k+\theta(t'))}\}, \eqno(6)$$ where
$E(k) =\sum_{m}{\langle0\vert{H_0}\vert{m}\rangle\exp(imk) }$ is the energy
dispersion
of the parent band.

To derive the eigenenergies, we note from (4) that the periodicity of
$u_{\epsilon
m}(t)$ implies that $C_m(T)=e^{-i[\epsilon T+ m\theta(T)]}C_m(0)$.
The Fourier transform of this relation yields the condition
$C_k(T)=C_{k+\theta(T)}(0)e^{-i\epsilon T}.$   Substituting (6) into this
condition yields that  $$C_{k+\theta(T)}(0)e^{-i\epsilon T}
=C_k(0)\exp\{-i\int_{0}^{T}{dt^\prime}{E(k+\theta(t'))}\}. \eqno(7)$$
The quasienergies can then be found from the condition that nonzero
solutions for the $C$'s exist.  In the pure AC case, we have $\theta(T)=0$,
showing that $k$ is a good quantum number.  The quasienergy is simply the
band energy $E(k+\theta(t'))$ averaged over one time period, whose $k$
dependence
can collapse into a constant under certain conditions [9].  Similar results
are obtained
for finite $F_0$ if the the matching ratio $F_0T/(2\pi)=\theta(T)/(2\pi)$ (note
that
$\hbar$ and the lattice constant are unity in our units) is an integer, because
$C_k$
is periodic in $k$ [10].

In the case of rational values $p/q$ of the matching ratio, where $p$ and $q$
are
relatively prime integers, $k$ is no longer a good quantum number.  However,
if we write the wave number in the form of $k=s+2\pi l/q$  where
$-\pi/q\le s< \pi/q$ and $l$  is an integer, then $s$ is conserved, which will
be used
to label the quasienergy states.  Since there are $q$ different $k$ states
coupled
together for a given $s$, we should obtain $q$ quasienergy bands.
We can obtain an analytical expression for these bands from the following
calculation.  We first substitute $k=s+2\pi l/q$ into both sides of
(7), and make a  product on each side over $l=0,..,q-1$, yielding
  $$\eqalign{
\prod_{l=0}^{q-1} C_{s+2\pi l/q+2\pi p/q}(0) &\times e^{-iq\epsilon T}\cr
=\prod_{l=0}^{q-1} C_{s+2\pi l/q}(0)&\times
\exp\{-i\int_{0}^{T}{dt^\prime}\sum_{l=0}^{q-1}{E(s+2\pi l/q+\theta(t'))}\} \cr
}
\eqno(8) $$
Next, since the $C$'s in the product all
have the same absolute value according to (7), a nontrivial solution to
(7) must correspond to a nonzero product of the $C$'s on each side of (8).
Finally, the products of $C$'s on the two sides are equal, because $C_k(0)$
is periodic with period $2\pi$, and because the set $\{0,...,q-1\}$ is the same
as
$\{p,...,q-1+p\}$ (mod. $q$).  Therefore, the exponentials on the two sides of
(8) must be equal, implying that the quasienergies must have the following form
$${\epsilon}_{ns} = {1\over qT}{\int_{0}^{qT}{dt\sum_{l=0}^{q-1} E(s+2\pi
l/q+\theta(t))}} + {2\pi{n} \over qT}, \hskip .5in  {(n= \rm  integer)}.
\eqno(9)$$
Since the quasienergy is defined within a Brillouin zone of width ${2\pi/T}$,
we only need to take $q$ consecutive values of $n$, e.g., $n=0,...,q-1.$
Finally, if we make a Fourier expansion of $E(s)$ in (9)
 and carry out the summation over $l$,  the  quasienergies then
have the simple form $$ {\epsilon}_{ns} =
\sum_{j}<0|H_0|qj>\int_{0}^{T}{dt\over T}
\exp\{iqj[s+\theta(t)]\}
 + {2\pi{n} \over qT}, \hskip .5in  {(n= \rm  integer)}.
\eqno(10)$$

Now let us consider a special DC--AC field of the form $F(t) =
F_{0}-F_{1}\cos(2\pi{t}/T)$. The vector potential (with a minus sign) is
then $\theta(t)=(p/q) [2\pi t/T-(F_1/ F_0)\sin(2\pi t/T)]$, where we have
used the rationality condition $F_0T/(2\pi)=p/q$.  The time integral in (10)
can
then be carried out in terms of the Bessel functions, yielding
$${\epsilon}_{ns}
=\sum_j <0|H_0|qj> J_{pj}({pjF_{1}\over F_{0}})\cos(qjs) +{n\over p}F_{0},
\hskip
.1in (n=\rm integer). \eqno(11)$$  Note that
the second term on the right hand side of (11) contains, in addition to the
integral Stark ladder $mF_{0}$ (when $ n/p = m$), the fractional ladder
structure
$nF_{0}/p$ (when $n/p\ne {\rm integer}$).  We call this additional feature the
dynamic fractional Stark ladder since the quasienergy gap between adjacent
Stark
levels is  $\Delta\epsilon = F_{0}/p$. A plot of
this quasienergy spectrum is presented in  Fig.1.

In order to generate the plot we have employed an exponentially  decaying form
for the matrix element to simulate the localized behavior of the Wannier
functions:
$<0|H_0|jq>={\Delta}_0\exp{-{\vert}jq/Q\vert},$ where ${\Delta}_0$
and $Q$ are constants associated with the lattice. The parameter values
utilized
in Fig.1 are given as follows: ${\Delta}_0=2.5meV$, $Q=5$, $F_0=2meV$,
 and $F_1=1meV$ (note that we have taken the lattice constant to be unity).
Along the
vertical axis are presented the inverse of the rational values of the matching
ratio
 ${F_0}T\over {2\pi}$ with denominators and numerators less than 30.  It
should be noted that we have chosen to fix the DC and AC field magnitudes;
therefore, the actual physical parameter being varied along the vertical
axis is the AC frequency. The quasienergy values label the horizontal
axis, and in order to obtain a symmetric graph, the plot extends over two
Brillouin zones of quasienergy, $[{-2\pi\over T},{2\pi\over T}]$, whose
width $2(q/p)F_0$ scales linearly with the vertical axis.
As it is the case that the quasienergy is taken over two Brillouin zones,
it is evident that for a given rational $p/q$ there will be $2q-1$
subbands, with subband centers evenly spaced with spacing ${F_0}/p$.  The
fact that a large denominator  $q$ of the matching ratio corresponds to a
small width of the quasienergy bands can be understood from (11).
Because of the  exponential decay of the hopping matrix element with
distance, we only need to  take the $j=\pm1$ terms for the dispersion of
the quasienergy band, yielding a bandwidth of $4\Delta_0 J_p({pF_1\over
F_0}) e^{-q/Q}$.  Apparent gaps in  the vertical direction surrounding
simple rational values are artifacts of the plot;  for a set of rationals
with denomenator less than a given integer, the density of points near a
rational of smaller denominator is just lower than that near  a rational
with a larger denominator.

Before attending to the important matter of experimental verification we will
briefly
discuss the  motivation underpinning the present investigation.
The time-dependent Schr{\"o}-dinger equation for a free electron is invariant
under
space--time translations. In the presence of a constant electric field, the
translation operators have to be modified by gauge transformations in order
to remain symmetry operations [12].  However, a translation in space no longer
commutes with that in time, unless ${\delta}x{\delta}t$ is an integral
multiple of $h/eE$, where $eE$ is the electric
force on an electron, and  ${\delta}x$ and ${\delta}t$ are the translation
step sizes in space and time
 respectively.  Here ${\delta}x$ is fixed to be an integer multiple of the
 crystal period $l$, and $1/{\delta}t$ is quantized in units of $eEl/h$,
which
 is just the frequency spacing of the Stark ladders. The fundamental
space--time
 area $h/eE$ is manifested in Wannier--Stark ladders.

Now let us introduce a periodic time variation to the system, so that we
have another fundamental space--time area $lT$ besides $h/eE$ of the
electric field. The matching ratio alluded to earlier, whose value
characterized the motion of carriers in DC--AC fields as being either localized
or
delocalized, can be expressed as the ratio of these two fundamental
space-time areas. This situation is not unlike that encountered in the
Hofstadter
problem, where subtle features appearing in the spectrum and wave functions
can be attributed to competition between the magnetic area $h/eB$ and the
unit cell area of a 2D periodic system [13]. The magnetic area arises in a
commutation condition  for the magnetic translation operators. Translations
in the x and y directions will commute if the product of the spatial step
sizes, ${\delta}x{\delta}y$, is an integral multiple of $h/eB$ where B is
the magnitude of a constant magnetic field oriented in the z direction [14].
Our
study should be an important step in understanding the congruence between
the electric and magnetic problems.

 In his paper, Hofstadter plotted the energy spectrum
vs. the rational values of a magnetic matching ratio between the
  magnetic area $h/eB$ and the
unit cell area of a 2D periodic system [13].
 A recursive  structure was manifest in his graph, and in our plot a degree of
self--similiarity is also evident.
However,  Hofstadter's spectrum has a skeleton structure of a `butterfly' in
which gaps appearing at rational matching ratios extend continuously in the
vertical direction [13], while in our spectrum the gaps at
rational matching ratios do not extend continuously in the vertical
direction.  The discrepency between the two spectra should not be too
disconcerting.  After all, in Hofstadter's plot it is the energy spectrum
which is investigated while our work focuses on a time dependent problem,
necessitating an investigation of the quasienergy. While the quasienergy is
in some sense a generalization of the energy, it might be more revealing to
look at the quasienergy from a different perspective. The quasienergy in
our problem originates in much the same way as the quasimomentum does in
spatially periodic systems. Adopting this view, it would be interesting to
revisit Hofstadter's problem. Now, however, fix the energy of the system and
consider a plot of one of the quasimomenta vs. the magnetic matching
ratio.  This issue  will be pursued further in the future.

In order to stimulate the experimental investigation of this fractional
ladder structure for the quasienergy spectrum of the systems discussed
above, consider the measurement of optical absorption in the following
problem. Suppose a weak laser field, described by a vector potential $\bf
A$, is incident on the system. This disturbance can be regarded as a
perturbation and accounted for by adding a term $-\bf{A\cdot P/\mu}$ to the
Hamiltonian (1). Here $\mu$ is the mass of an electron, and $\bf P$ is the
momentum operator. Subject to the condition ${\theta}(T)/(2\pi) =p/q$ the
Floquet states for the unperturbed Hamiltonian (1),
 ${\psi}_{ns}(x,t) ={\exp}[-i{\epsilon}_{ns}t]u_{ns}(x,t)$,  provide a
useful basis for perturbative calculations. Wave functions of the perturbed
system can then be expanded in terms of the functions  ${\psi}_{ns}(x,t)$:
${\Psi}(x,t) = \sum_{ns}a_{ns}(t){\psi}_{ns}(x,t)$. Thus, consider the
system to initially be in a particular Floquet state
${\psi}_{n^{\prime}s}^i(x,t)$ before the perturbation is switched on. As a
standard radiation theory approximation,  we neglect the momentum of the
photon and choose ${\bf A} = A_{0}\sin({\omega t})\hat x$. If we
distinguish our final state as ${\psi}_{ns}^{f}(x,t)$, the
probability transition rate is then given by

$${{\rho}_{fi}(NT)\over NT}=2\pi\sum_l \delta(\epsilon_{ns}^f
-\epsilon_{n^\prime s}^{i}\pm \omega -l2\pi/T)
|{A_0\over 2\mu}
\int_{0}^{T} {dt\over T} P_{nn^\prime}(t) \exp\{il 2\pi t/T
\}{\vert}^2. \eqno(12)$$
where $P_{nn^\prime}(t)$ is the optical matrix element, defined by
$P_{nn^\prime}(t) = \int{dxu_{ns}^{f*}(x,t)Pu_{n^\prime s}^{i}(x,t)}$ [9].
The existence of absorption peaks at nonzero $l$ is analogous to the Umklapp
process in a crystal where the quasimomentum is conserved upto a reciprocal
lattice vector.

 Note that had we specified that the initial and final states came from
quasienergy bands generated by the splitting of the same parent band, the
frequencies at which absorption occured would simply be $\omega =\pm (n -
n^\prime){F_{0}/p}$ upto integer multiples of $2\pi/T$. Such an example is
somewhat
artificial in that it is not plausible experimentally, because it is not even
clear
how one can arrange for one quasienergy band to be  fully occupied while
another is  completely empty.

   A more realistic example can be derived
from the system used by the authors [6,7] in their study of the usual
Wannier-Stark ladders:  a semiconductor superlattice such as that of
GaAs-AlGaAs in which only  the highest miniband in the occupied valence
band and the lowest miniband in the empty conduction band are involved in
the optical transition.  In the presence  of a DC electric field, these
minibands are turned into the usual Wannier-Stark ladders, so the optical
transitions are really between such ladder states.  In our case, the
application of an AC electric field in addition to a DC field should turn
these minibands into quasienergy bands with the fractional ladder
structures as discussed above.  We have numerically calculated the optical
absorption spectrum in  such a system, using an approximation that the
initial miniband has a negligiable  band width, so that the spectrum
actually reflects the quasienergy structure of the final miniband.  The
final miniband is
modeled as $E(k)=2\Delta_0 \cos(k)$, which gives zero width quasienergy
bands for all but integer values of the matching ratio (corresponding to
q=1, see Eq.(11)). In Fig.
2a, the spectrum for the DC-only case is presented.  The DC field is taken as
${\pi \over 2}\Delta_0$ per
lattice constant (which is unity in our units).  Note that the highest peak
is at $\omega=E_g$, the gap  between the center of the initial and final
minibands.  Weaker spectral lines appear at frequency intervals of $F_0$,
revealing the usual Wannier-Stark ladder structure. In fig. 2b, the
spectrum with an additional AC field with a frequency of ${2\over 5} F_0$
($p=5$, $q=2$) and a strength of $F_1=0.3F_0$ is presented.  It is noted
that additional spectral lines appear at intervals of $F_0/5$, revealing the
fractional ladder structure in the quasienergy spectrum. A full
calculation of the optical absorption spectrum with realistic experimental
parameters will be presented elsewhere.

Physically, we can look at the absorption problem as a system of Stark
ladders driven by a strong AC field at frequency $\Omega$ and perturbed by
a weak AC field at frequency $\omega$. Electrons may make transitions
between the integer ladder states by simultaneous absorption of a photon at
$\omega$ and $l$ photons at $\Omega$, where $l$ can be negative
corresponding to the case of emission. Multiple absorption or emission of
photons is possible because the AC field of frequency $\Omega$ is not
weak. From  energy conservation we have the relation ${E_g}+{kF_0} =
\omega + l\Omega$. This shows that $\omega - E_g$ can be any integer
combination of $F_0$ and $\Omega$. For the case when $F_0 = (p/q)\Omega$,
the resulting absorption lines occur in intervals of ${\Omega}/q =
{F_0}/p$, where $p$ and $q$ are relatively prime integers.

To conclude we emphasize the following points.
 The dynamic fractional Stark ladder arises out of a competition between
two fundamental space--time areas, one associated with the DC field and the
other with the AC field. Hence the simultaneous presence of both types of
fields is essential. When a temporally
periodic electric field interacts with a spatially periodic system, a
parent band will split into a series of quasienergy subbands if the
matching ratio
is a rational number $p/q$.   There are $q$ such subbands if the
quasienergy is
restricted to a Brillouin zone of length $2\pi/T$.
 If only the DC field is present, the Hamiltonian (1)
becomes time independent, and therefore the `quasienergy' is exactly the
energy. In that case the fractional Stark ladder changes into the integral
Stark ladder.

 In our treatment of optical absorption, the effect of electron
scattering by phonons and impurities has been neglected. This approximation is
valid
provided that the relaxation time for scattering is long compared to the period
of
Bloch oscillations. If such a condition cannot be met, the role of electron
scattering should be involved in the description of electron dynamics and
optical absorption.

\vskip .5 in  We are greatful to S. -G.
Chen, Z. -B. Su, Y. Avron, and S. Fishman for many useful discussions.  We
also extend our thanks to J. Liu for suggesting this
collaboration.  This work
was supported  by the NNSF of China
and the  Chinese Academy of Science. Additional
support was provided by the NIST and the Robert A. Welch  Foundation.

\vskip .5 in {\bf References} \vskip .2 in
\item {1.} G. H. Wannier, Phys.
Rev. {\bf117}, 432 (1960); Rev.Mod.Phys. {\bf 34}, 645 (1962).
\item {2}  J. E. Avron, J. Zak, A. Grossmann, and L. Gunther, J. Math. Phys.,
{\bf 18,}
918 (1977).
\item {3} J. B. Krieger and G. J. Iafrate, Phys. Rev. B {\bf 33}, 5494 (1986).
\item {4.}
R. W. Koss and L .N. Lambert, Phys. Rev. B {\bf 5}, 1479 (1972).
\item {5.}
L. Esaki and R. Tsu, IBM J. Res. Dev. {\bf 14}, 61 (1970).
\item {6.} J.
Bleuse, G. Bastard, and P. Voisin, Phys. Rev. Lett. {\bf 60}, 220 (1988).
\item {7.} A wealth of experimental reports appear in the literature ; for
example, E. E. Mendez, F. Agull{\'o}--Rueda, and J. M. Hong, Phys. Rev.
Lett. {\bf 60}, 2426 (1988); P. Voisin, J. Bleuse, C. Bouche, S. Gaillard, C.
Alibert, and A. Regreny, Phys. Rev. Lett. {\bf 61}, 1639 (1988); H. Schneider,
K.
Fujiwara, H.T. Grahn, K. v. Klitzing, and K. Ploog, Appl. Phys. Lett. {\bf
56}, 605 (1990); and H. Schneider, A. Fischer, and K. Ploog, Phys. Rev. B
{\bf 45}, 6329 (1992).
\item {8.} D. H. Dunlap and V. M. Kenkre, Phys. Rev.
B {\bf 34}, 3625 (1986).
\item {9.} M. Holthaus, Phys. Rev. Lett. {\bf 69},
351 (1992);  and Z. Phys. B {\bf 89} 251 (1992); M. Holthaus and D. Hone,
Phys. Rev. B {\bf 47}, 6499, and {\bf 48}, 15128 (1993).
\item {10.} J. Zak, Phys. Rev. Lett. {\bf 71},
2623 (1993). \item {11.} X. -G. Zhao, Phys. Lett. A {\bf155}, 299 (1991); {\bf
167},
291 (1992).  \item {12.} Q. Niu,
Phys. Rev. B {\bf 40}, 3625 (1989); N.Ashby and S.C.Miller, Phys. Rev. {\bf
139},
A428 (1965).
\item {13.}  D. R. Hofstadter, Phys. Rev. B {\bf14}, 2239 (1976).
\item {14.}  E. Brown, Phys. Rev. {\bf 133}, A1038 (1964); J. Zak, ibid.,
{\bf134},
A1607 (1964).
\vfill\eject

Caption for Fig.1:

Plot of the quasienergy vs. the inverse rational values, $q/p$, of the matching
ratio
 $-F_{0}T/2\pi$.  For each $q/p$ the quasienergy is taken over two Brillouin
zones,
$[-2{\pi}/T, 2{\pi}/T]$.  The parameters in Eq. (11) are set at ${\Delta}_{0}$
= 2.5meV,
$F_{0}$ =2meV, $F_{1}$ = 1meV and Q = 5. (Only rationals with $q < 30$ and $p <
30$ are shown.)

\vskip

Caption for Fig.2:

Plot of the optical absorption strength I (in arbitrary units) as a
function  of the optical frequency $\omega$. (a) The DC-only case, with
$F_0={\pi\over2}\Delta_0$.  (b) With an additional AC field of frequency
${2\over 5}F_0$ and strength $F_1=0.3F_0$.

 \bye